\documentclass[12pt,preprint]{aastex}

\shorttitle{Electron Shock Surfing Acceleration}
\shortauthors{Hoshino and Shimada}

\received{2002 January 22}
\begin{document}

\title{ Nonthermal Electrons at High Mach Number Shocks: \\
   Electron Shock Surfing Acceleration}

\author{M. Hoshino and N. Shimada}
\affil{Department of Earth and Planetary Physics, University of Tokyo, Tokyo, Japan}
\email{hoshino@eps.s.u-tokyo.ac.jp}

\begin{abstract}
We study the suprathermal electron acceleration mechanism in a perpendicular magnetosonic shock wave in a high Mach number regime by using a particle-in-cell simulation.  We find that shock surfing/surfatron acceleration producing the suprathermal electrons occurs in the shock transition region where a series of large amplitude electrostatic solitary waves (ESWs) are excited by Buneman instability under the interaction between the reflected ions and the incoming electrons.  It is shown that the electrons are likely to be trapped by ESWs, and during the trapping phase they can be effectively accelerated by the shock motional/convection electric field.  We discuss that suprathermal electrons can be accelerated up to $m_i c^2 (v_0/c)$, where $m_i c^2$ is the ion rest mass energy and $v_0$ is the shock upstream flow velocity.  Furthermore, some of these suprathermal electrons may be effectively trapped for infinitely long time when Alfv\'en Mach number $M_A$ exceeds several 10, and they are accelerated up to the shock potential energy determined by the global shock size.
\end{abstract}


\section{Introduction}

The origin of high energy particles seen radiating in astrophysical synchrotron sources is still a major unresolved problem of high-energy and plasma astrophysics.  These sources include supernova remnants (SNRs) and extragalactic radio sources by jets etc. \citep[e.g.,][]{Koy95,Mei89,Car96,Jun99}.  The shock acceleration has been discussed as one of important processes producing the high-energy particles, and there are many theoretical and observational efforts on the high-energy particle acceleration/heating so far.  The diffusive/Fermi acceleration process particularly met with great success  \citep{Axf77,Bel78,Bla78}, because it predicts a power-law energy spectrum with the power-law index of 2 that depends weakly on the shock compression ratio, and because the index suggests a very close to that believed to exist in many astrophysical sources.  

From {\it in situ} measurements at the planetary shocks and at the interplanetary shocks, the energetic ions with a power-law spectrum has been observed  \citep[e.g.,][]{Sch80,Gos81}, and they are believed to be produced by the diffusive shock acceleration.  For the energetic electrons in the astrophysics context, the radio spectra of SNRs and the emission from active galactic outflow, for example, are believed to be produced by the shock acceleration, and much work has been concerned with the diffusive shock acceleration of electrons.  However, the energetic electrons with a power-law spectrum at the interplanetary shocks and at the planetary bow shocks are rarely ever observed \citep[e.g.,][]{Shi99}, and the diffusive-type acceleration seems to be malfunctioning in the Heliosphere.  The electron nonthermal acceleration still remains an unresolved issue of considerable interest.

In order to the shock diffusive acceleration to start, a certain pre-acceleration mechanism should provide a seed population having from thermal to mildly high energy, and then the energized particles can be subsequently accelerated to highly relativistic energies.  Our understanding of the electron pre-acceleration mechanism, which is the so-called injection process, is very limited \citep[e.g.,][]{Lev96}.  Furthermore, MHD waves which provide the resonant scattering of electrons in the shock upstream are not well studied so far.

By taking into account plasma instabilities in detail, \citet{Pap88,Car88} proposed the electron energization process at high Mach number shocks, in which electron heating is produced through two-step instabilities in the shock transition resion where the reflected ions coexist with the incident ions and electrons \citep{Ler82,Wu84}.  Buneman instability (BI) \citep{Bun58} is first excited by the velocity difference between the reflected ions and the incident electrons, and the electrons are heated up by the instability.  As the next step, the ion acoustic (IA) instability is triggered under the pre-heated electron plasma by BI, and the electron temperature increases by a factor approximately equal to $M_{A}^{2}$/$\beta_{e}$, where $M_A$ is Alfv\'en Mach number and $\beta_e$ is the upstream electron plasma beta.  This scenario of the electron heating mechanism has been demonstrated by using a hybrid simulation code where ions are treated as particle but electrons are assumed to be massless fluid \citep{Car88}.  Recently \citet{Non00} extended their study to the shock structure including the electron dynamics by using a particle-in-cell simulation where both ions and electron are treated as particle.  They found the formation of the phase space ``electron hole" in association with a localized, large-amplitude, electrostatic solitary wave in the shock front region.  It is discussed that BI in the nonlinear stage produces a large-amplitude, localized electrostatic wave which is called as the electrostatic solitary wave (ESW) \citep{Dav70,Omura94}.  They also found that the ESW itself is almost stable, but the interaction with other waves in the inhomogeneous shock transition region leads to the dynamical evolution of the shock energy dissipation.  They discussed that the electrostatic solitary wave (ESW) plays an important role not only on the rapid electron thermalization but also on the nonthermal electron acceleration, and that the ESWs obtained by the simulation can be compared with {\it in situ} observations of the electric field waveform at the Earth's bow shock \citep{Mat97,Bal98}.  However, the actual physics of the nonthermal electron acceleration processes has remained elusive.

In this paper, we study in details the suprathermal electron acceleration through the interaction of electrons with ESWs in a transverse, magnetosonic, high Mach number shock.  We discuss that the electrons trapped by ESW can resonate with the shock motional electric field, and the so-called shock surfing mechanism is effective for producing the non-thermal, high-energy electrons \citep{Hos01,McC01}.  Shock surfing acceleration at quasi-perpendicular shocks is usually considered for ions to be a pre-acceleration mechanism to initiate diffusive shock acceleration \citep{Sag66,Sag73,Sug79,Kat83,Lem84,Ohs85,Zan96,Lee96,Uce01}.  The electrostatic field at the shock front directs upstream, and the ions which energy cannot overcome the electrostatic potential well are reflected upstream in a super-critical shock of $M_A > 2.7$.  In the standard shock surfing, the ions are trapped between the shock front and the upstream by the Lorentz force.  During the reflection process, ions travel along the shock front and can be accelerated by the motional/convection electric field.  In this mechanism, however, electrons can be neither reflected nor accelerated.  We discuss that a series of large amplitude electrostatic waves excited by the nonlinear Buneman instability in the shock transition region can play a role of counterpart of the shock front potential for the case of the ion shock surfing, and those ESWs can effectively trap electrons.  Then the electron shock surfing mechanism can be switched on. 

We also propose that for a large Mach number shock with $M_A >$ several 10, electrons are likely to be trapped for infinitely long times, because the electrostatic force becomes always larger than the Lorentz force.   In this situation, the electrons can be perfectly trapped by ESWs, and can be quickly gain their energies until they obtain the global transverse shock potential energy.  We discuss that the shock surfing/surfatron near the shock front region is the efficient accelerator for the relativistic electrons.

\section{Nonthermal Electrons in High Mach Number Shocks}

We study the ion-electron dynamics organizing the electron energization and wave activities in the shock transition region by using the one-dimensional, particle-in-cell simulation code where both ions and electrons are treated as particle \citep{Hos92}.  In our simulation study, a low-entropy, high-speed plasma consisting of electrons and ions is injected from the left boundary region which travels towards positive $x$.  At the injection boundary at $x=0$, the plasma carries a uniform magnetic field $B_z$, polarized transverse to the flow.  
The downstream right boundary condition is a wall where particles and waves are reflected.  Under the interaction between the plasma traveling towards positive $x$ and the reflected particles and waves from the right-hand boundary, the shock wave is produced, and it propagates backward in the $-x$ direction.  Then the downstream bulk plasma speed becomes zero in the simulation frame.  The grid size is comparable to the electron Debye length in downstream, includes about 1800 particles for each species in downstream.
The plasma parameters are as follows; upstream plasma $\beta_{e} = \beta_{i}$ = 0.01 ($\beta_{j} = 8{\pi}nT_{j}/B^2$, where $n$, $T$, and $B$ are, respectively, the density, temperature, and magnetic field strength), the ratio of the plasma frequency to the electron cyclotron frequency $\omega_{pe}/\omega_{ce}=19$ ($\omega_{pe} = \sqrt{4 \pi ne^2/m_e}$, $\omega_{ce} = eB/(m_ec)$), and the ratio of ion to electron mass $m_i/m_e$ = 20.  The whole system size $L/(c/\omega_{pe})=300$, the upstream speed $v_0$ in the simulation frame is $0.25c$, and the ratio of Alfv\'en speed to the speed of light is $V_A/c = 1.2 \times 10^{-2}$.  The shock Alfv\'en Mach number becomes $M_A \sim 32$ in the shock frame.  (Note that the upstream speed in the shock frame is about $(3/8)c$.)  Our simulation is performed until $t\omega_{pe} = 1815$, when the shock front reached at $X/(c/\omega_{pe}) \sim 100$.

Figure 1 shows a snapshot of the nonlinear stage of a collisionless shock front region at $t \omega_{pe} = 735$.  The left-hand panel shows the shock front region from $X/(c/\omega_{pe}) = 150$ to $300$, while the right-hand panel is its enlarged picture.  The leftward and the rightward regions are respectively the shock upstream and downstream.  From the top, the ion phase space diagram in (X, $U_{ix}$), the electron phase space diagrams in (X, $U_{ex}$), and in (X, $|U_e|$) where $|U_e| = (U_{ex}^2 + U_{ey}^2 + U_{ez}^2)^{1/2}$.  The bottom three panels are the transverse electric field $E_y$, the magnetic field $B_z$, and the electric field $E_x$.  The plasma four velocity $U$ is normalized by the upstream flow velocity $U_0 = v_0 /\sqrt{1-(v_0/c)^2}$.  The magnetic field, and the electric field are normalized by the upstream magnetic field $B_0$ and the upstream motional electric field $E_0 = v_0 B_0/c$, respectively.  The spatial scale is normalized by the electron inertia length $c/\omega_{pe}$ in upstream.  

In the shock front region in Figure 1, we can find two ion components in the ion phase space diagram; one is the cold ion flowing into the shock downstream, the other is the reflected ion going away from the shock front.  This ion dynamics has been well studied by both the satellite observations \citep{Pas81,Gos82,Sck83} and the theory/simulation studies for super-critical Mach number shocks \citep{Ler82}.  A part of the incoming ions is reflected due to both the polarization electrostatic field and the compressed magnetic field in shock downstream.  The magnetic overshoot structure can be found around $X/(c/\omega_{pe}) \sim 248$.

In addition to such ion dynamics around this shock transition region, we find the electron hole in the electron phase space diagram in $(X, U_{ex})$ around $X/(c/\omega_{pe}) \sim 223$, and its corresponding electrostatic solitary wave (ESW) in the electric field $E_x$.  Other ESWs in the growing phase can be also found in the foreside of the largest ESW structure.  \citet{Non00} discussed that those localized structures are excited by Buneman instability between the reflected ions and the incoming electrons \citep{Pap88,Car88,Die00a,Die00b}.  The simulation result in Figure 1 is basically same as that obtained by \citet{Non00} and \citet{Schm01}.

Looking at the electron hole structure in the shock transition region in details, we find that some electrons are accelerated and gain a large amount of energy.  After the strong acceleration in the shock transition region, the electrons are transported downstream.
Figure 2 shows the downstream electron energy spectrum. The energy $\gamma = \sqrt{1 + (U_e/c)^2}$ is normalized by the incident electron bulk energy $\gamma_0 = 1/\sqrt{1 - (v_0/c)^2} \sim 1.03$.  The spectrum is superposed over 538 snapshots for the time interval from $\omega_{pe}t=535$ to $1815$. 
The dot-dashed line is the Maxwellian fit for the spectrum, and the bottom dotted line is the so-called one-count level, which is the reciprocal of the number of snapshots of $1/538$ in this case.  Although the statistics of the nonthermal high energy tail is not good, but we can clearly find the enhancement of the nonthermal population above the Maxwellian level over $\gamma/\gamma_0 > 1.8$.  

\citet{Non00} also obtained that the similar suprathermal energy spectrum for the high Mach number shock of $M_A \sim 10$.  In addition, they found that the suprathermal electrons are not generated for a low Mach number case of $M_A \sim 3$, and the spectrum is well fitted by a thermal Maxwellian.  With regard to the electron energy spectrum obtained by these simulations, a high Mach number shock seems to be a candidate for producing high energy electrons.

Before discussing the mechanism producing the suprathermal electrons, we would like to give a comment that the suprathermal electron energy is much larger than the potential energy of ESW.  The potential energy $e \phi_{\rm esw} = e E_{\rm esw} \Delta_{\rm esw}$ can be estimated as
$20 (\omega_{ce}/\omega_{pe})(v_0/c) m_e c^2 \sim 0.26 m_e c^2$, 
where we have assumed $E_{\rm esw} \sim 20 E_0$, the width of ESW, $\Delta_{\rm esw} \sim c/\omega_{pe}$ from Figure 1.  Other simulation parameters are $\omega_{ce}/\omega_{pe} = 1/19$ and $v_0/c = 1/4$.  Then we get the increment of the Lorentz factor normalized by the initial Lorentz factor as $\Delta \gamma/\gamma_0 \sim (\gamma - \gamma_0)/\gamma_0 \sim 0.26$.  Therefore the electrons cannot be accelerated over $\gamma/\gamma_0 \sim 1.26$.  Moreover, due to a bipolar electric field structure of ESW, no energy gain is expected after crossing the whole ESW in the x-direction.

\section{Electron ``Shock Surfing Acceleration''}

In order to know how and where the high energy electron is accelerated in a high Mach number shock, we study the electron trajectory near the shock front region.  Figure 3 shows the stack plot of the $E_x$ field and two typical electron orbits (left-hand panel) and the time history of the total momentum $|P_e| = m_e |U_e|$ (right-hand panel).  The total momentum $|P_e|$ is normalized by the upstream momentum $P_0 = m_e U_0$.  The vertical axis $t$ is normalized by the electron plasma frequency $\omega_{pe}$, and the time history is plotted after $t_0 \omega_{pe} = 535$.  The most of waveforms in the shock transition region are basically propagating the negative $x$ direction, which speeds are slightly faster then the shock front propagation speed in the wave growing phase, because ESWs are generated by the velocity difference between the reflected ions and the incoming electrons, and because the reflected ions carry larger momentum than the incoming electrons.  As the wave amplitude increases in the nonlinear phase, the speed of ESW becomes slow down, and the waveform is convecting downstream.  We find that some waves are sporadically transported downstream during $(t-t_0)\omega_{pe} = 200 \sim 420$.  Note that the downstream plasma is at rest in our simulation, and the shock front is propagating towards the negative $x$ direction.

Let us first look at the trajectory denoted by the solid line (left-hand panel).  The particle is situated in the upstream region before $(t-t_0)\omega_{pe} \sim 100$, and it is convected towards the shock front with the $E \times B$ convection speed of $v_0 = c E_0/B_0$.   When the particle arrived at the shock front at $(t-t_0)\omega_{pe} \sim 120$, it is trapped by a large amplitude ESW until $(t-t_0)\omega_{pe} \sim 220$, and after that it escapes out of the ESW structure and is convecting downstream.  From the time history diagram (right-hand panel), we find that the efficient energization occurs while the electron is trapped for the time interval from $(t-t_0)\omega_{pe} = 120$ to $240$.  The gradual energy increase can be also found after the detrapped phase from $(t - t_0)\omega_{pe} = 240$ to $500$.  The second phase of acceleration corresponds to the region having the magnetic field gradient in association with the magnetic overshoot.

The dashed line shows another typical electron orbit for which the interaction of electron with ESW was very weak.  When crossing the shock front region, the electron was not trapped by ESW, and it passed through ESW without any significant energy gain.  The sinusoidal oscillation means the gyro-motion of particles.  Note that the solid curve has a larger sinusoidal oscillation than the dashed curve, because the electron denoted by the solid curve gains larger energy during the shock front crossing.

Figure 4 shows the time history of the first adiabatic invariant of $P_{\perp e}/B^2$ for the same particles analyzed in Figure 3.  $\langle P_{\perp e}/B^2 \rangle$ is averaged over the upstream electron gyro-period of $\omega_{ce}^{-1}$, and is also normalized by the upstream value of $P_{\perp 0}/B_0^2$.  The dashed curve is almost constant, and we know that the electron motion is almost adiabatic during the shock crossing.  For the solid curve case that had the strong interaction of electron with ESW, we find a sharp energy increase for the time interval from $(t-t_0)\omega_{pe} = 120$ to $240$, which is suggestive a non-adiabatic acceleration.  After $(t-t_0)\omega_{pe} \sim 240$ we found the gradual energy increase in Figure 3b, but the baseline of the normalized invariant seems to be almost constant except for a sinusoidal oscillation.  This suggests that the electron heating is almost adiabatic after the electrons are convected downstream, and the non-adiabatic acceleration occurs only when the electron interacts with ESW in the shock transition region.

Figure 5 shows the above two particle trajectories in the $x-y$ plane.  Both two trajectories are drawn from $(X,Y) \sim (200~c/\omega_{pe},0)$ at $\omega_{pe}t = 535$ in the upstream region, and the electrons are convected towards positive $x$ along $y=0$.  The Larmor radii are very small in the upstream region because their thermal velocities are very small.  Note that this result was obtained by the one-dimensional, particle-in-cell simulation, but both positions of $x$ and $y$ are calculated.  Also note that the $x$ and $y$ distances are plotted on the different scale in Figure 5.  They encounter the shock front region around $X/(c/\omega_{pe}) = 225$.  The notation of the solid and dashed curve is same as that in Figures 3 and 4.  We find that the electron denoted by the solid curve are traveling along the $y$ axis in the shock front region, it can be understood that the main energy gain of the electrons comes from the motional electric field.  In fact, the energy gain of electron from the motional electric field $\Delta \varepsilon_m$ can be estimated as 
$e E_m \Delta Y$, 
where $\Delta Y \sim 60 (c/\omega_{pe})$ and $E_m = (v_0 - v_{\rm esw}) B_0 /c$ is the motional electric field in the ESW frame.  Since the propagation speed of ESW is almost same as the speed of the reflected ions \citep{Non00}, we may assume $v_{\rm esw} \sim -v_0$.  Then we get $\Delta \varepsilon_m \sim 120 (\omega_{ce}/\omega_{pe})(v_0/c) m_e c^2 \sim 1.6 m_e c^2$, and the Lorentz factor of the accelerated electron becomes 2.6.  Namely, the normalized momentum $P_e/P_0 \sim 9$, which is almost consistent with the momentum gain in Figure 3.

From the above analysis of the simulation data, we think the so called ``shock surfing'' mechanism plays an important role on electron acceleration.  Let us quickly review the idea of the shock surfing \citep{Sag66,Sag73,Sug79}.  The shock surfing mechanism has been extensively studied for the ion acceleration \citep{Kat83,Lem84,Ohs85,Zan96,Lee96,Uce01}.  Due to the inertia difference between ions and electrons flowing into the shock, the polarization electric field normal to the shock front is formed.  An ion having a small velocity normal to the shock front, i.e., $1/2 M v_x^2 < e \phi_s$, can be reflected from the shock front to the shock upstream, where $\phi_s$ is the electrostatic shock front potential induced by the inertia difference between ions and electrons.  During the gyro-motion in upstream, ions gain their energy from the motional electric field parallel to the shock front.  
If the width of the shock front potential $\phi_s$ is thin, the electric force can overcome the Lorentz force, and then the multiple reflection can occurs.  In a quasi-perpendicular shock, \citet{Osh87} studied the nonlinear steepening of magnetosonic waves, and showed that if $\omega_{pe}/\omega_{ce} < 1$, the thickness of the shock front potential $\phi_s$ becomes of the order of the electron inertia scale $c/\omega_{pe}$.  They then demonstrated by using a particle simulation the strong resonant interaction of the reflected ions with the motional electric fields.  

What we discuss here is the ``electron'' shock surfing acceleration \citep{Hos01,McC01}.  The above ``standard'' shock surfing acceleration cannot apply for the electron acceleration, because the electron cannot be reflected from the shock front by the shock front potential $\phi_s$.  We propose a new scheme of ``electron'' shock surfing acceleration under the action of ESW.  Since the electron hole is a positively charged structure, and an electron can be trapped if $m_e v_x^2/2 < e \phi_{\rm esw}$, where $\phi_{\rm esw}$ is the scalar potential for the electrostatic solitary wave (ESW).  (Note that the Lorentz force is also an important agent for the particle trapping, but we will discuss this point later.)  Furthermore, the propagation velocity of ESW differs from the plasma bulk velocity, and ESW together with the trapped electrons can stay longer time in the shock transition region.  Namely, in the frame moving with ESW, the convection electric field is not zero.  Therefore, we think that the so called ``shock surfing'' mechanism is occurring for electrons.  

Figure 6 summarizes our idea of the electron shock surfing mechanism.  Top panel shows an electron's trajectory in the $x-y$ plane.  The magnetic field is polarized perpendicular to the $x-y$ plane, and the plasmas are convecting towards positive $x$.  The shock upstream is the left-hand side, while the downstream is the right-hand side.  Bottom panel shows the electric field $E_x$ along the $x$ axis, and ESW in association with its electron hole in phase space is depicted in the center.  Due to the nature of the electron hole, the electron charge density is slightly lower than the ion one, and ESW has a bipolar signature with diverging electric field.  If an electron convecting towards the ESW structure is reflected by both the Lorentz force and the electric field $E_x$ and is trapped inside the ESW structure, it is successively accelerated towards the negative $E_y$ direction.   
As increasing the electron's velocity $v_y$ by the shock surfing/surfatron acceleration, it can be de-trapped from ESW when the Lorentz force $e v_y B_z/c$ becomes larger than the electric force $e E_x$, and then it is convecting towards downstream and becomes quickly an isotropic, gyrotropic distribution.  The time variable nature of the ESW evolution may be the other important factor to control the de-trapping process.

\section{Unlimited Electron Acceleration}

To illustrate the efficiency of the above electron shock surfing mechanism, let us discuss the $x$ component of the equation of motion,
\begin{equation}
  \frac{d}{dt} P_x = e E_x + \frac{e}{c} v_y B_z,
\end{equation}
where $P_x = m_e U_{ex}$ is the electron momentum.  In our interest, $E_x$ on the right-hand side can be replaced by the electrostatic solitary wave (ESW) produced in the shock transition region.  If the electric force of $e E_{\rm esw}$ is larger than the Lorentz force of $(e/c) v_y B_0$, the electrons are trapped and gain their energy.  During the non-adiabatic acceleration phase, the velocity $v_y$ increases.  If the electron satisfies the condition of $e E_{\rm esw} < (e/c) v_y B_0$, it can escape from ESW and will be convected downstream.

The amplitude of ESW may be estimated by equating the wave energy density to the drift energy density of the incoming electron \citep{Ish81}, and we obtain,
\begin{equation}
   \frac{E_{\rm esw}^2}{8 \pi} \sim \frac{1}{2} m_e n V_d^2 \alpha,
\label{eq:eswamp}
\end{equation}
where $V_d \sim 2 v_0$ is the relative velocity between the reflected ions and the incoming electrons.  The energy conversion factor $\alpha$ is of order of $O(1)$.  Although the nonlinear saturation process of BI still remains controversial, we adapt here the conversion factor $\alpha$ discussed by \citet{Ish81,Die00b}, and we use,
\begin{equation}
   \alpha = \frac{1}{4} \sim (\frac{m_e}{m_i})^{1/3}.
\end{equation}

Before discussing the efficiency of electron acceleration, it is better to check whether or not the above estimation is reasonable.  We first analyze the magnetospheric observation data for a high Mach number shock.  From the satellite observations in the Earth's bow shock, we know that $E_{\rm esw} = 50 \sim 200~{\rm mV/m}$ \citep{Mat97,Bal98}, while the solar wind, motional electric field $E_0$ is a few mV/m.  Then, the ratio of $E_{\rm esw}/E_0$ is of order of $10^1 \sim 10^2$.  On the other hand, by virtue of Eq.(\ref{eq:eswamp}), the ratio of $E_{\rm esw}$ to the motional electric field $E_0 = v_0 B_0/c$ in upstream can be obtained,
\begin{equation}
  \frac{E_{\rm esw}}{E_0} = 2 \frac{c}{V_A} \sqrt{\alpha \frac{m_e}{m_i}}.
\label{eq:esw}
\end{equation}
Since the Alfv\'en velocity $V_A$ is about 300 km/s in the solar wind, the theoretical estimation of $E_{\rm esw}/E_0$ becomes $13 \sim 23$.  This value is remarkably close to the observation.   In the same way as this, we can check this for our simulation result.  We observed $E_{\rm esw}/E_0 = 10 \sim 30$ in Figure 1, while Eq.(\ref{eq:esw}) suggests $E_{\rm esw}/E_0 = 19 \sim 23$ for $m_i/m_e = 20$ and $V_A/c = 0.012$.  The comparison between the theoretical estimation and the simulation gives a good agreement again.  Therefore, we think the estimation of the amplitude of ESW in Eq.(\ref{eq:eswamp}) is reasonably well.

Now we can estimate the maximum electron energy.  The electric force $F_E = e E_x$ can be given by,
\begin{equation}
  F_E \sim e E_{\rm esw} 
\sim 2 \left( \frac{v_0}{V_A} \right) e B_0 \sqrt{\alpha \frac{m_e}{m_i}},
\end{equation}
and the Lorentz force $F_B$ in the shock transition region is,
\begin{equation}
  F_B \sim \frac{e}{c} v_y B_0.
\end{equation}
By equating $F_E$ to $F_B$, we obtain the maximum velocity for $v_{\rm y,max}$ as,

\begin{equation}
v_{\rm y,max} \sim
\left\{
\begin{array}{ll}
 2 c M_A \sqrt{\alpha \frac{m_e}{m_i}} & 
\mbox{for $2 M_A \sqrt{\alpha (m_e/m_i)} < 1$} \\
c & \mbox{for $2 M_A \sqrt{\alpha (m_e/m_i)} \ge 1$} 
\end{array}
\right.
\end{equation}

\noindent
If $M_A > (1/2) \sqrt{m_i/(m_e \alpha)}$, then $F_E > F_B$ is always satisfied, and the maximum speed becomes the speed of light.  In this case, the electrons cannot escape from the ESW region, and the electrons continue to accelerate until they reach the edge of a global shock structure.  The condition of the ``unlimited electron acceleration'' for the real mass ratio of $m_i/m_e = 1830$ is given by,
\begin{equation}
   M_A > 43 \sim 75.
\end{equation}

\noindent
We think that the electron surfing with ESW in the shock transition region is a strong accelerator.

\section{Conclusions and Discussion}

We discussed the electron shock surfing through a large amplitude, small-scale electrostatic field in the shock transition region.  The electrons may be unlimitedly accelerated when Alfv\'en Mach number $M_A > 43 \sim 75$.   It is well known that the dynamical shock emerges for a super-critical shock with Alfv\'en  Mach number $M_A > 2.7$, and a fraction of incoming ions are reflected from the shock front.  Due to the velocity difference between the incoming electrons and the reflected ions, a series of large amplitude electrostatic waves are excited by the nonlinear Buneman instability.   We found that the electrons can be trapped by the large amplitude wave and the non-adiabatic acceleration of electron occurs under the interaction with the shock motional electric field parallel to the shock front.  We also found that, as further increasing the Mach number $M_A$ and if $M_A$ exceeds $43 \sim 75$, the electrons can be unlimitedly accelerated and gain the relativistic energy.  Such high Mach number shocks can be found in SNRs, which are known as the relativistic electron accelerator with a strong X-ray emission.

It is an open question whether or not the shock surfing acceleration discussed in this paper can produce the power-law energy spectrum.  For the case of the ion shock surfing, \citet{Zan96} pointed out that shock surfing can produce the non-thermal power-law energy spectrum in the shock front region.  It may be interesting to study the similar scheme for the electron shock surfing.  Our understanding on the electron shock surfing is still limited, but we would like to emphasize that the acceleration rate of the shock surfing is much faster than that of the shock diffusive/Fermi acceleration, and it may be important not only for the injection process of the shock diffusive acceleration but also for high energy particle acceleration.

The unlimited electron shock surfing acceleration is based on the assumption that ESW stably exists during the interaction.  It is known, however, that the shock front is not stable and periodically breaks.  Namely, the shock reforms with a cycle time of ion gyro-poeriod, $\omega_{ci}$, which is called as the shock reformation process \citep{Qu86}.  The reflected ion, which can excite ESWs through Buneman instability, has also the same dynamical behavior with the shock reformation, and BI may be quenched every shock reformation cycle.  Therefore, most of electrons may be de-trapped from ESW in association with the shock reformation process.  Assuming that the ion reformation time scale is about ion gyro-period, $\omega_{ci}^{-1}$, the maximum energy of the electron shock surfing might be limited by,
\begin{equation}
   \varepsilon_{\rm max} \sim \frac{e E_0 c}{\omega_{ci}} \\
    \sim m_i c^2 (\frac{v_0}{c}).
\label{eq:emax}
\end{equation}

\noindent
In the case of the simulation parameter used in Figure 2, the above maximum energy normalized by the upstream flow energy can be estimated as $\gamma_{\rm max}/\gamma_1 \sim (m_i/m_e)(v_0/c) \sim 7.5$.  However, the electron energy density became below the one-count level around $\gamma/\gamma_0 \sim 4.5$ due to the limited number of particles in the simulation box.  It seems that we have to increase the number of simulation particle at least 100 times more to examine the cut-off signature, and we need the larger computer resources to understand the behavior of high energy spectrum.

Our simulation was performed in one-dimensional system, but the multi-dimensional effect may be important for the non-thermal particle acceleration.  One may think that ESW has a finite transverse scale in two- or three-dimensional system, and that the finite ESW structure may limit the electron shock surfing acceleration.  On the other hand, it may be possible that some electrons move from an ESW to another ESW in a stochastic manner and resonate successively with the motional electric field, because the shock surface is expected to have a wavy structure.  In this case, the maximum energy of the shock surfing is not limited by Eq.(\ref{eq:emax}), and we think that the electrons can be unlimitedly accelerated until they reach the edge of a global shock structure.  We are planning to study two-dimensional shock dynamics as the future problem.

The propagation speed of the electrostatic solitary wave against the shock front is also an important agent to control the electron surfing acceleration.  If ESW is moving with the same speed of the plasma medium, the shock motional electric field $E_m$ in the ESW frame disappears, and no surfing acceleration occurs.  In fact, we find that the efficient acceleration occurs in the upstream side of the shock transition region where the amplitude of ESW is not necessarily large.  ESWs are growing their amplitudes towards downstream in the transition region, but the ESW structure starts to move with the almost same velocity with the plasma medium, and the motional electric field in the ESW frame becomes small.  Then, the electrons cease the surfing acceleration.  However, ESWs start to collapse by colliding with other waves in inhomogeneous shock transition region \citep{Non00}, and those collapsed waves may provide further stochastic and surfatron acceleration, as discussed recently by \citet{McC01}.

In this report, we restrict our attention to the weakly magnetized shock where the electron cyclotron frequency $\omega_{ce}$ is smaller than the electron plasma frequency $\omega_{pe}$, because most astrophysical plasmas have a weakly magnetized plasma with $\omega_{ce} < \omega_{pe}$.  In the case of $\omega_{ce} \gg \omega_{pe}$, Buneman instability and its subsequent nonlinear evolution discussed in this paper may be suppressed \citep{Die00a}. \citet{Bes99} studied a quasi-perpendicular magnetosonic shock with $\omega_{ce} \gg \omega_{pe}$ using a particle-in-cell simulation code, and investigated acceleration of electrons during the nonlinear decay process of the magnetosonic wave.  It is interesting to know the efficiency of suprathermal electron acceleration as the function of $\omega_{ce}/\omega_{pe}$.

Finally, we should like to make the point that it may be possible to explore many aspects of the electron shock surfing acceleration from {\it in situ} observations at the interplanetary shocks and at the planetary bow shocks.  It is known that the interplanetary shocks initiated by the solar flare may develop into a high Mach number shock with $M_A \sim 40$ around the Mercury orbit of $\sim 0.4$ AU \citep[e.g.,][]{Smart85}.  Such observations could bridge a gap of our understanding between the electron injection and acceleration at the interplanetary/planetary bow shocks with relatively low Mach numbers and the astrophysical shocks with very high Mach numbers.

\acknowledgments
The authors are grateful to T. Terasawa for fruitful discussion.  MH acknowledges support from the International Space Science Institute (ISSI) at Bern/Switzerland for the collisionless shock working group.

\clearpage

\begin{figure}
\plottwo{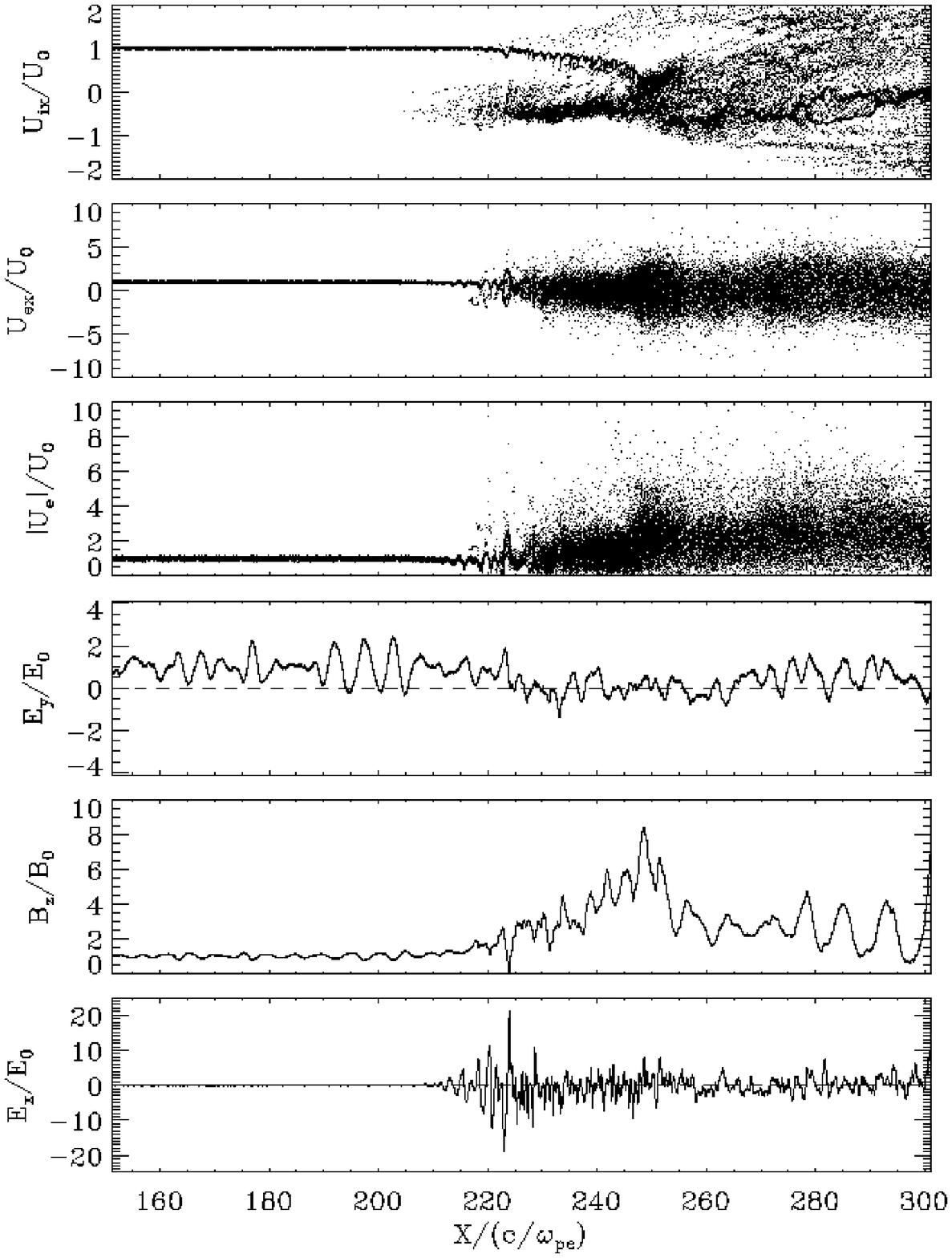}{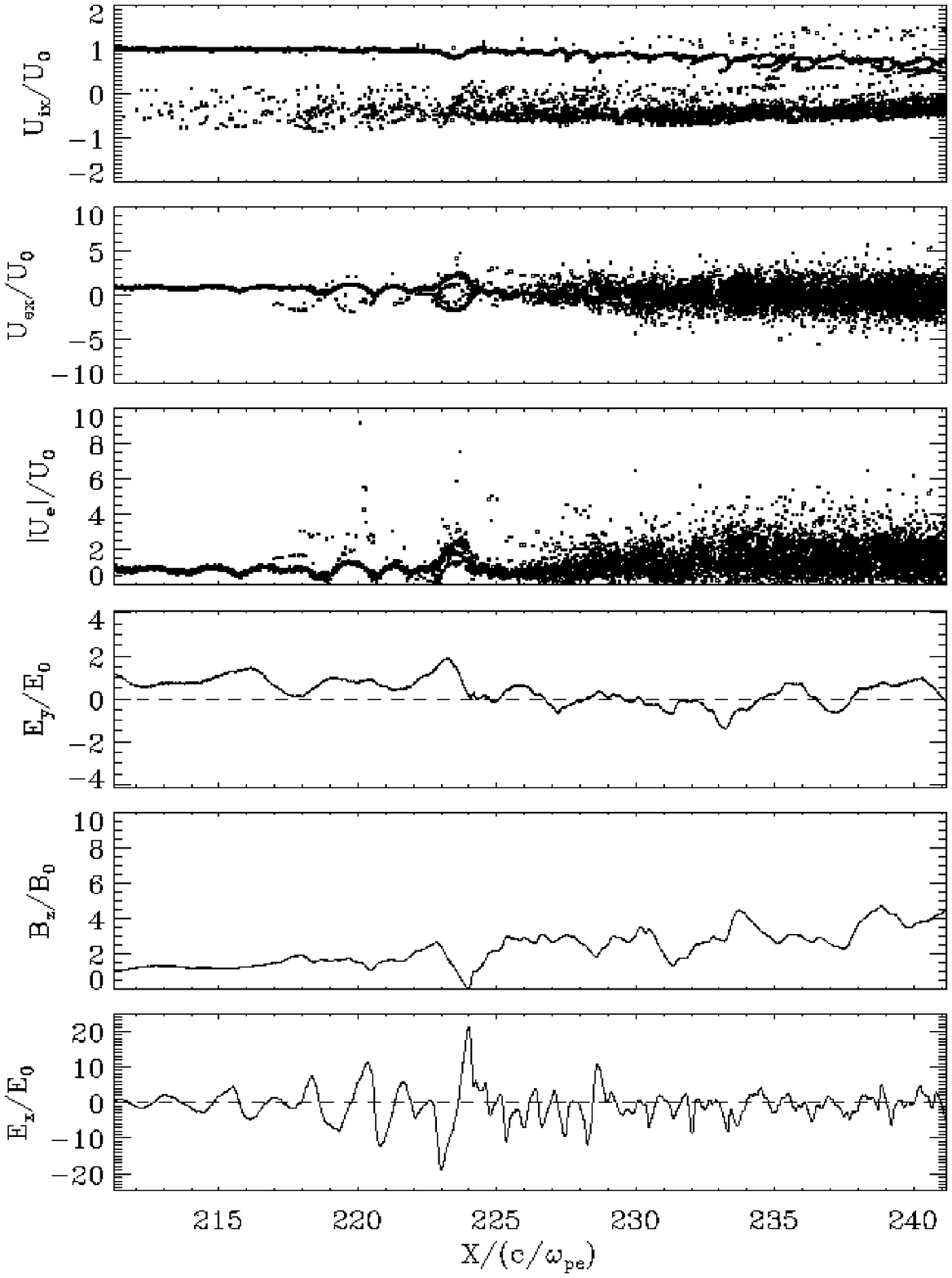}
\caption[fig1.eps]{High Mach number, magnetosonic shock structure for $M_{A} = 32$: an overall shock front region (left), and an enlarged picture in the shock transition region (right).
\label{fig1}}
\end{figure}
\clearpage

\begin{figure}
\plotone{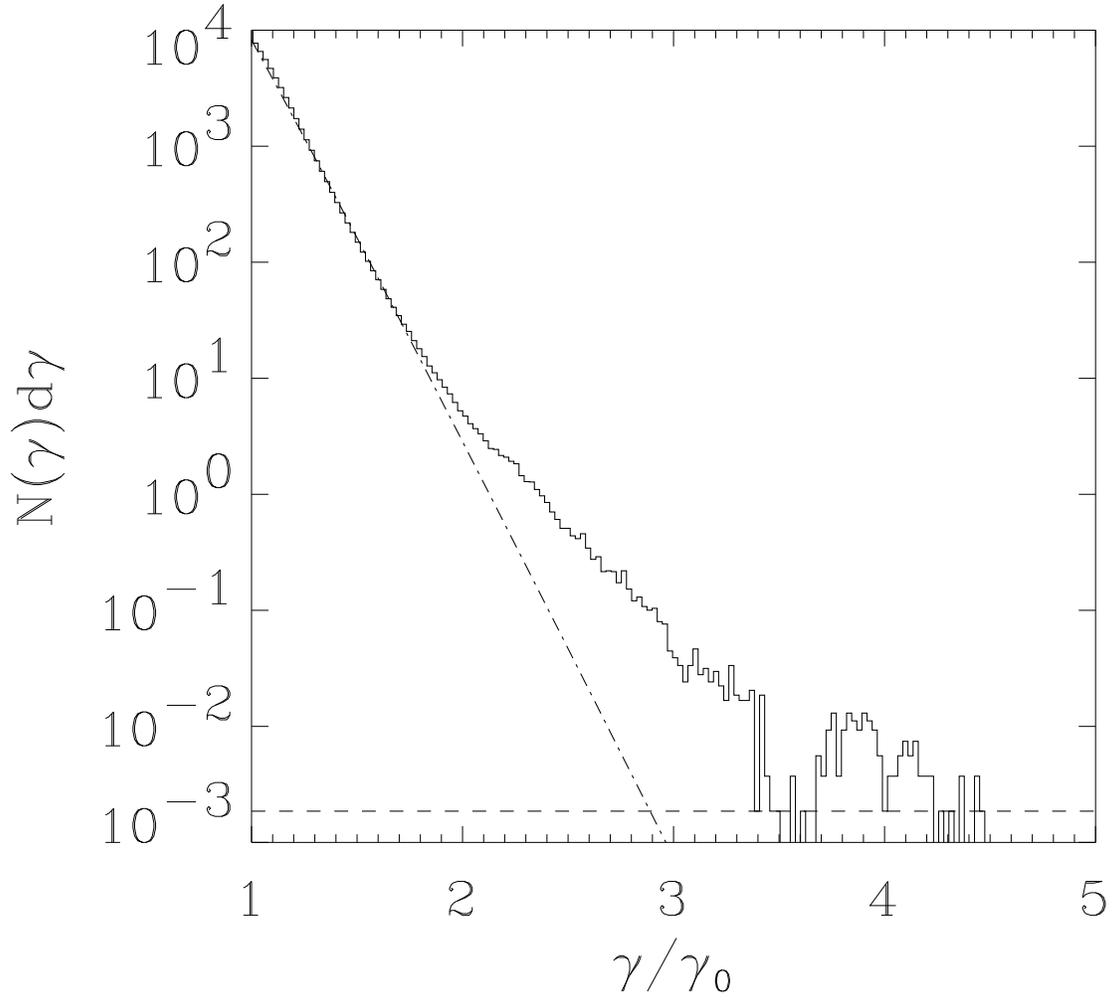}
\caption{ Downstream suprathermal electron energy spectrum. Dotted lines indicate Maxwellian distribution.
\label{fig2}}
\end{figure}
\clearpage

\begin{figure}
\plottwo{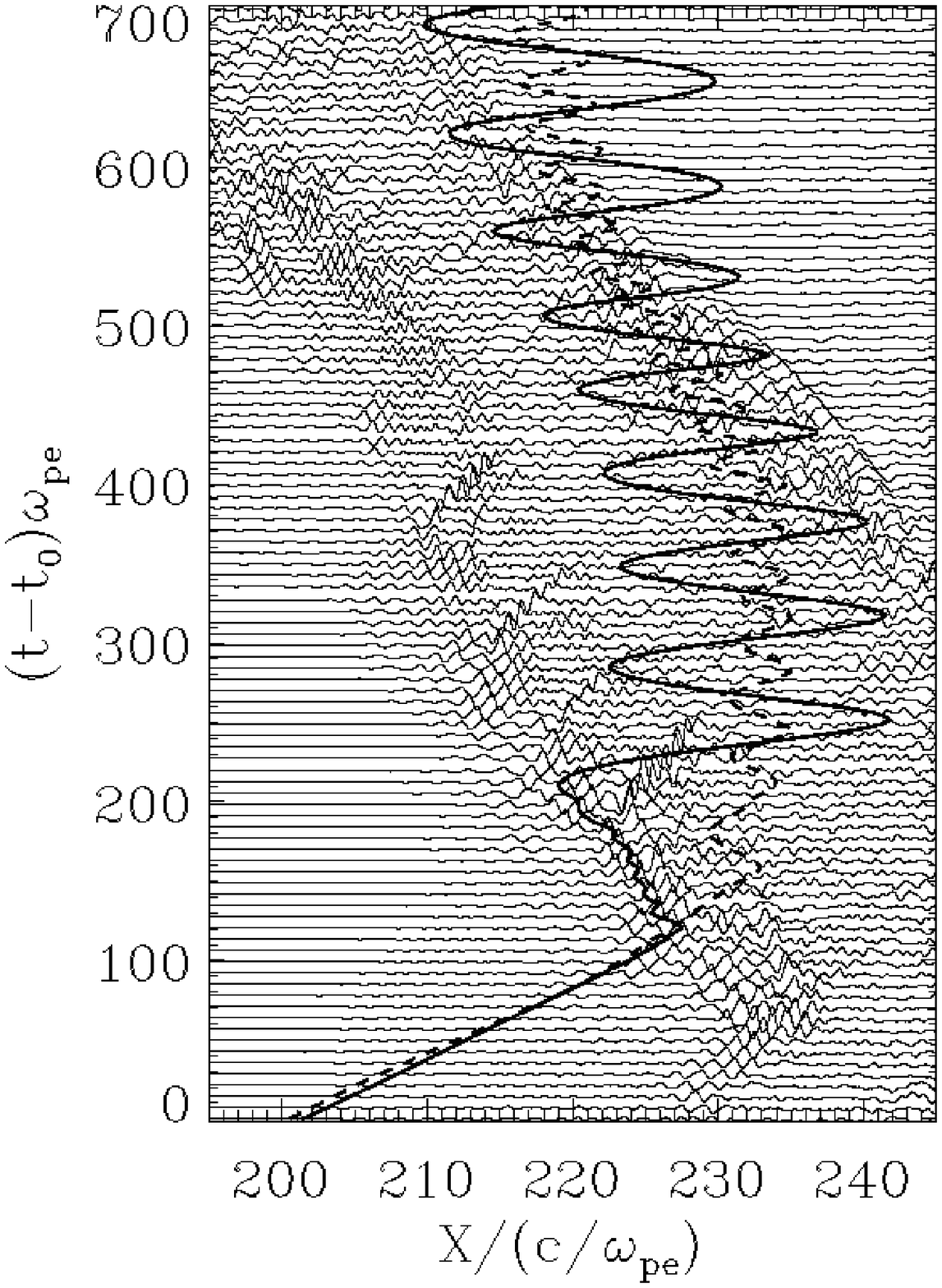}{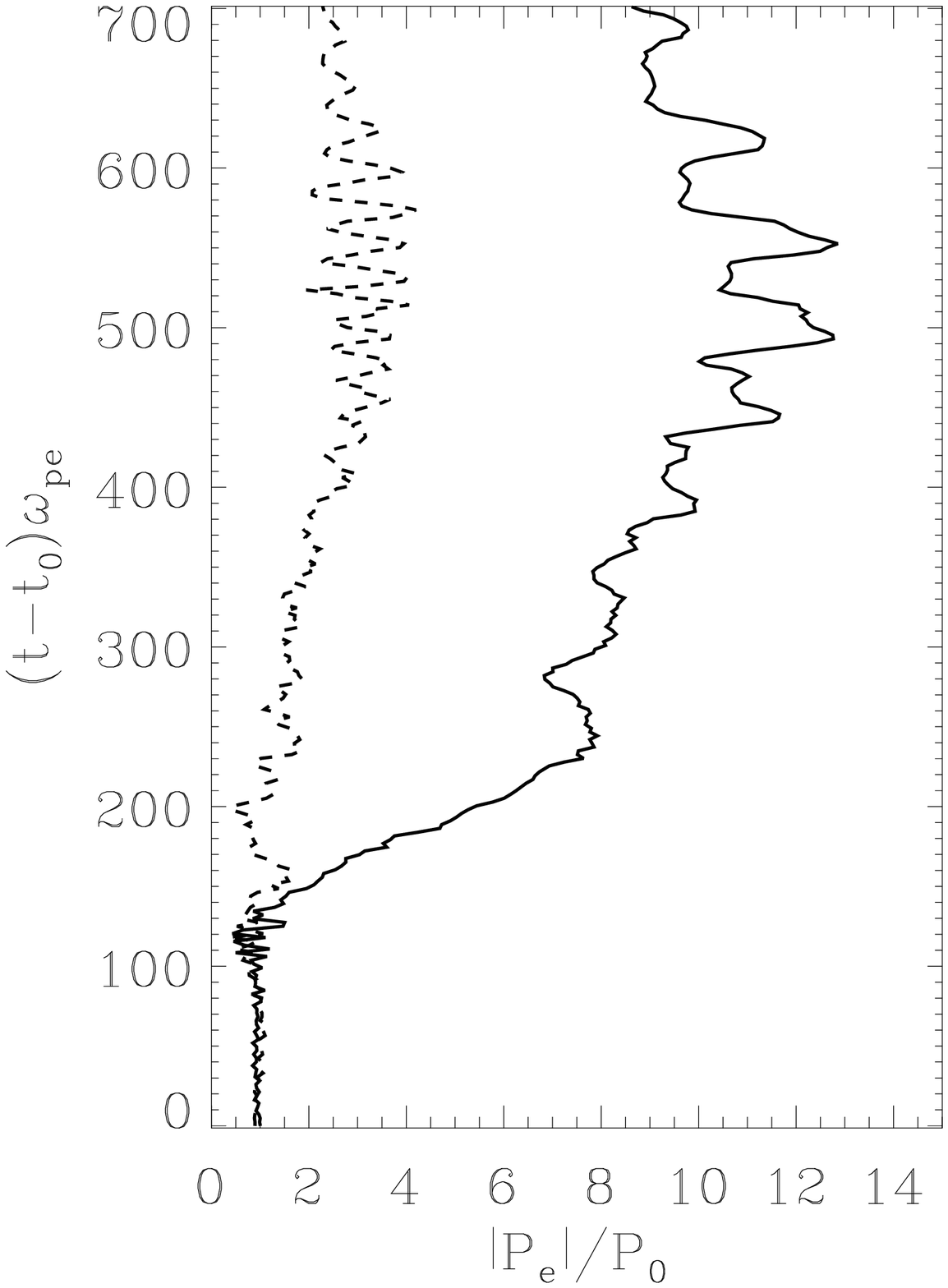}
\caption{Time evolution of wave form for the longitudinal electric field $E_x$ and two typical particle trajectories (left), the time history of the total momentum (right).
\label{fig3}}
\end{figure}
\clearpage

\begin{figure}
\plotone{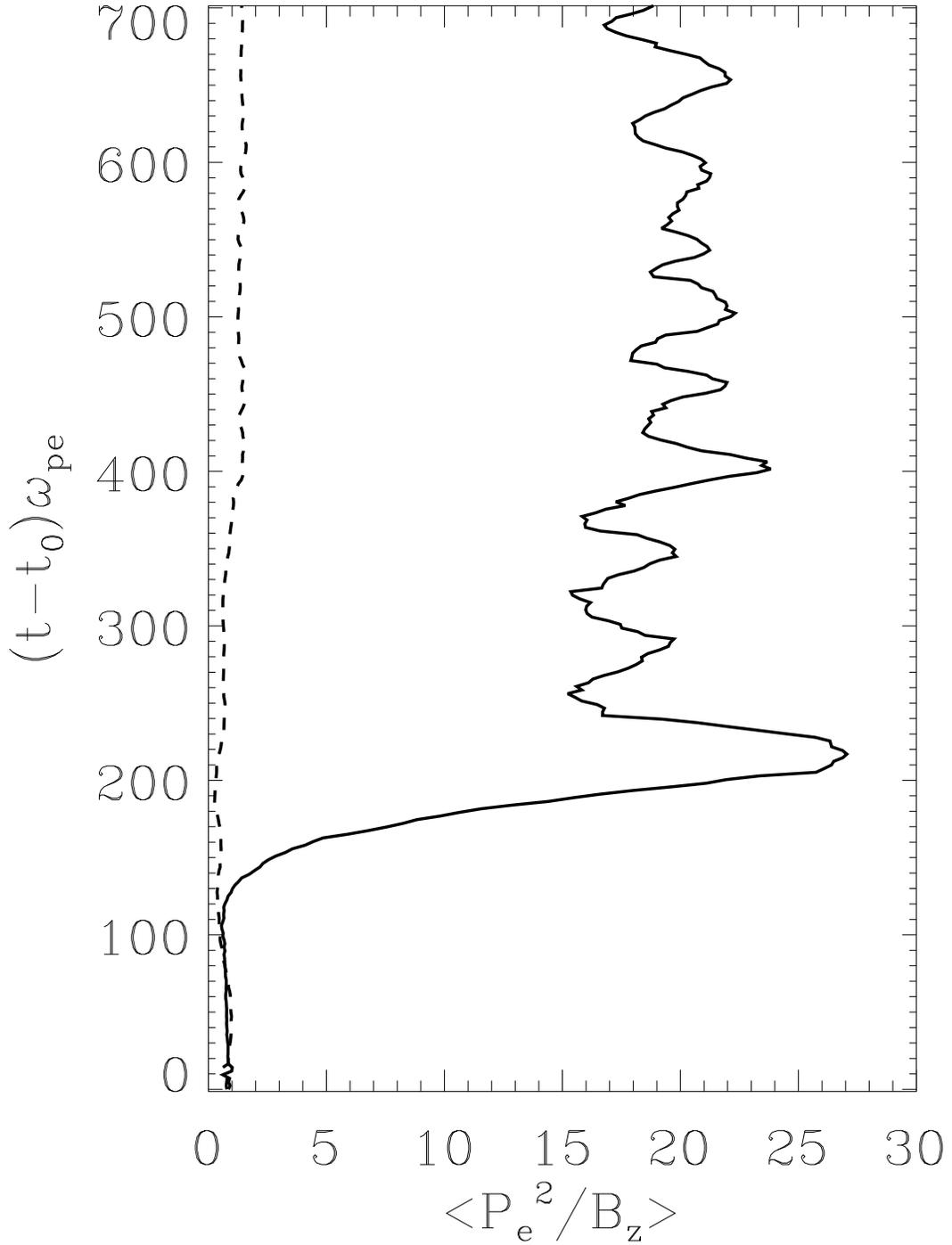}
\caption{The time history of the 1st adiabatic invariant $P_e/B^2$ for the two typical particles shown in Figure 3. 
\label{fig4}}
\end{figure}
\clearpage 

\begin{figure}
\plotone{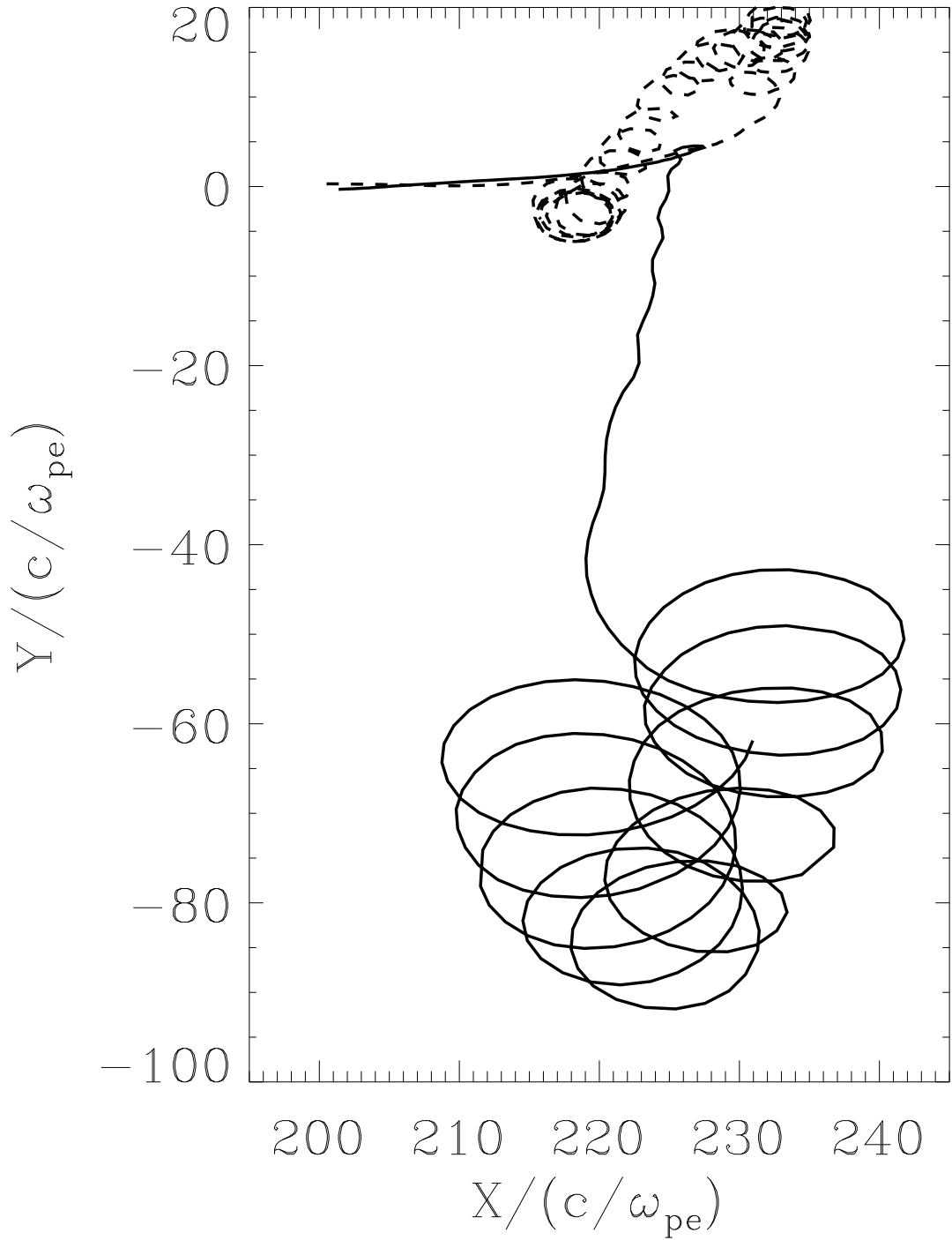}
\caption{The particles trajectories in the plane perpendicular to the magnetic field $B_z$ for the two typical particles shown in Figure 3. 
\label{fig5}}
\end{figure}
\clearpage 

\begin{figure}
\plotone{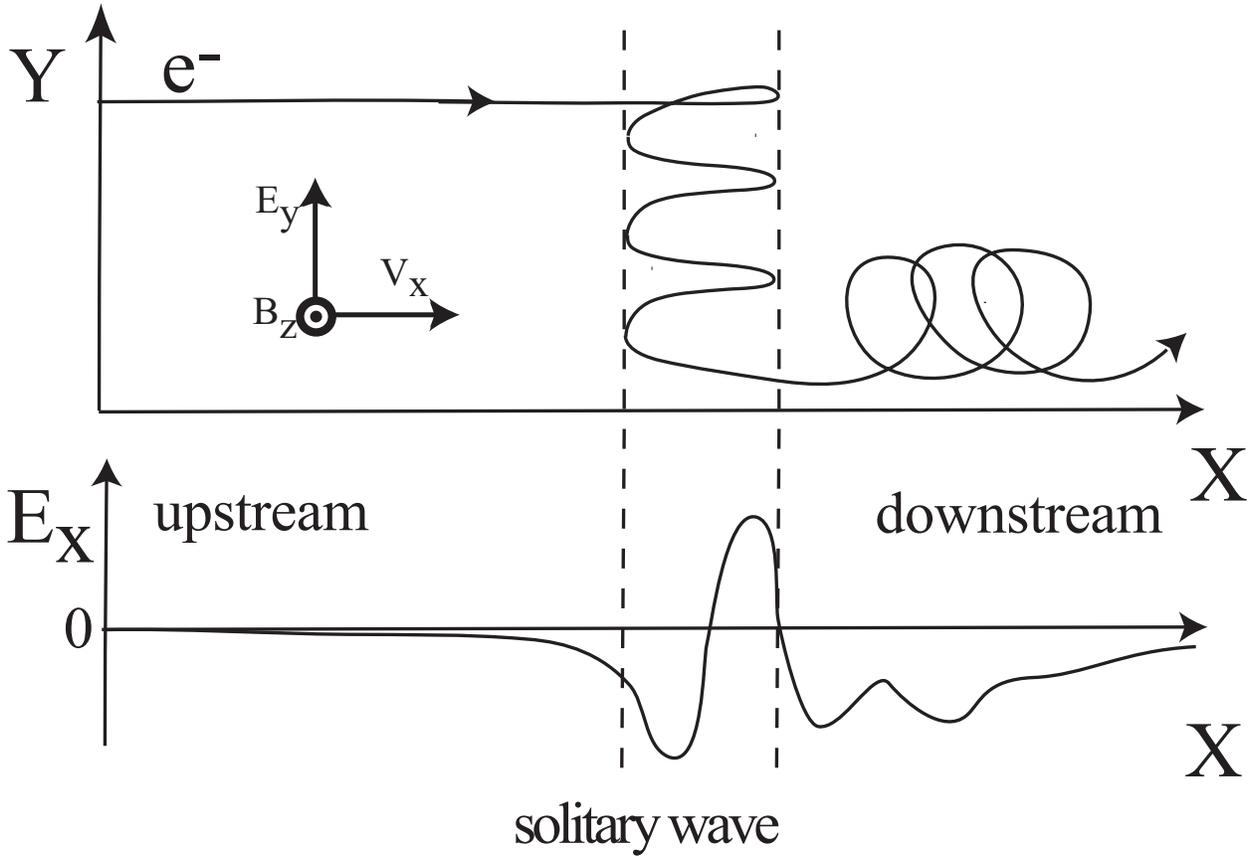}
\caption{An illustration of the electron shock surfing mechanism under an action of ESW (electrostatic solitary wave). 
\label{fig6}}
\end{figure}
\clearpage 

\end{document}